\documentclass[%
 reprint,
superscriptaddress,
nofootinbib,
 amsmath,amssymb,
prc,
floatfix,
]{revtex4-2}

\usepackage{graphicx}% Include figure files
\usepackage{dcolumn}% Align table columns on decimal point
\usepackage{bm}% bold math
\usepackage[pdftex,colorlinks,linkcolor = blue,urlcolor  = blue,citecolor = blue]{hyperref}
\usepackage{siunitx}
\DeclareSIUnit\noop{\relax}

\begin{document}

\preprint{APS/123-QED}

\title{Impact of level densities and $\gamma$-strength functions on $r$-process simulations}

\author{F.~Pogliano}
\email{francesco.pogliano@fys.uio.no}
\affiliation{Department of Physics, University of Oslo, N-0316 Oslo, Norway}

\author{A.~C.~Larsen}
\email{a.c.larsen@fys.uio.no}
\affiliation{Department of Physics, University of Oslo, N-0316 Oslo, Norway}

\date{\today}% It is always \today, today,

\begin{abstract}

Studies attempting to quantify the sensitivity of the $r$-process abundances to nuclear input have to cope with the fact that the theoretical models they rely on, rarely come with confidence intervals. 
This problem has been dealt with by either estimating these intervals and propagating them statistically to the final abundances using %one-zone 
reaction networks within simplified astrophysical models, or by running more realistic astrophysical simulations using different nuclear-physics models consistently for all the involved nuclei. 
Both of these approaches have their strengths and weaknesses.
In this work, we run $r$-process calculations for five trajectories using 49 different neutron-capture rate models.
Our results shed light on the importance of taking into account shell effects and pairing correlations in the network calculations.
\end{abstract}

\maketitle

%\tableofcontents

%\section{\label{sec:level1}First-level heading:\protect\\ The line
%break was forced \lowercase{via} \textbackslash\textbackslash}

\section{Introduction}

The origin of elements heavier than iron in the solar system %may be explained by 
can be attributed to three main production mechanisms, the $p$, the $s$ and the $r$ processes~\cite{B2HF, Cameron57}. 
While the first one is only responsible for the creation of some few, proton-rich nuclei, the latter two are each responsible for about 50\% of the final solar abundances. 
The $r$ process stands for \textit{rapid} neutron-capture process, involves very high neutron densities ($\gtrsim 10^{20}$ cm$^{-3}$) and the neutron flux lasts for less than a second. 
Within the timescale of the $r$ process, the neutron-capture rate is usually higher than  the $\beta$-decay rate, making it possible to reach very exotic and neutron-rich nuclei up to the neutron drip line. During the $r$ process, material will pile up at $A \approx 80$, 130 and 195 due to the high neutron flux as well as the neutron closed shells at $N = 50, 82, 126$. 
The $r$-process sites have been a mystery for many years. 
The recent observation of the GW170817 gravitational waves event~\cite{GW170817} from a neutron star merger (NSM) followed by its electromagnetic counterpart (e.g., Ref.~\cite{Arcavi2017}) provided observational evidence that heavy elements indeed are produced in these cataclysmic astrophysical events. 
Many other sites have been proposed over the years, such as prompt core-collapse supernovae, collapsars and the reheating of supernova ejecta by neutrinos (neutrino-driven wind, NDW), see e.g.~Ref.~\cite{Kajino2019} for a review.

The $r$ process may happen under different astrophysical conditions and can be categorized as either hot or cold. The hot $r$ process happens in environments with temperatures above a few GK. 
When temperatures exceed $\approx 2.0$~GK the photo-disintegration rate wins over the $\beta$-decay rate and one reaches an $(n,\gamma)\leftrightharpoons(\gamma,n)$ equilibrium, where only temperature, neutron abundance and irradiation time, the neutron separation energy $S_n$ of the involved nuclei, their $\beta$-delayed neutron emissions and $\beta$ decay rates are important for the final abundances~\cite{ArnouldGoriely2007}. 
If the $r$ process instead happens mostly at colder temperatures, then other nuclear properties such as %lifetimes, branching ratios, 
fission properties and neutron-capture reaction rates will also be crucial for the nucleosynthesis flow.

In order to model the $r$ process in a realistic way, we need both the correct astrophysical conditions and the right nuclear input~\cite{Pinedo2023}. 
Radioactive-beam facilities and new experimental techniques may allow us to study the properties of some of the involved, exotic nuclei such as masses, $\beta$-decay rates and $(n,\gamma)$-rates (see e.g.~the reviews in Refs.~\cite{Mumpower2016, *Mumpower2016e, PPNP_larsen} and references therein).
Nevertheless, simulations are still heavily reliant on the predictions of theoretical models.

Sensitivity studies are a tool that can help us figure out the properties of which nuclei or nuclear regions have the biggest impact on the final abundances, and are thus of vital importance for planning and conducting  experiments. 
These studies usually focus on one specific quantity (such as $\beta$-decay rates, masses or neutron-capture rates) by varying this quantity within some confidence interval or range of uncertainty, and then analyze the impact this variation has on the final abundances~\cite{ArnouldGoriely2007, Mumpower2016, *Mumpower2016e}. 
Although all of these quantities are important, in this work we will  focus on the neutron-capture rates.

The lack of confidence intervals in theoretical neutron-capture rates predictions poses a great problem when trying to estimate the uncertainties in the final abundances stemming from the uncertainties in the nuclear models. 
In addition to the need for realistic simulations of the astrophysical environment(s), this lack is posing a significant challenge for sensitivity studies in the attempt to nail down which nuclei or regions in the nuclear chart that influence the $r$-process abundances the most. 

Several sensitivity studies for neutron-capture rates can be found in the literature. 
These studies may be divided into two categories, depending on the approach used to address these astrophysical and nuclear aspects of the uncertainties. 
We label the first category the ``statistical approach''. 
Here, a particular set of initial conditions for the $r$ process is chosen so that the network simulation reproduces some known abundances (e.g. the solar $r$-process abundances, or the ones from a more sophisticated astrophysical simulation). 
The  abundance yields from this simulation are referred as the ``baseline''. 
The neutron-capture rates of a set of the involved nuclei are then modified individually, and for every rate variation the $r$-process simulation is run again with the same initial conditions, and every new output is compared to the baseline. 
The nuclei whose variation will lead to the biggest change in abundances with respect to the baseline, are identified as the most interesting to study experimentally. 
The capture rates are varied either by multiplying and dividing the baseline rates by a constant factor~\cite{Surman2009, Mumpower2012, Surman2014, Vescovi2022}, by assuming the rate uncertainty to have a log-normal distribution about the baseline value~\cite{Mumpower2016, *Mumpower2016e} or by assuming the distribution to be flat within the predictions of a set of theoretical models, and log-normal outside~\cite{Nikas2020}. 
In the last two cases, the probability distributions are used to pick the variations using a Monte Carlo technique, and the results are then analyzed statistically~\cite{Mumpower2016, *Mumpower2016e, Mumpower2015c, Mumpower2015c2}. 

\begin{figure}
\includegraphics[width=0.50\textwidth]{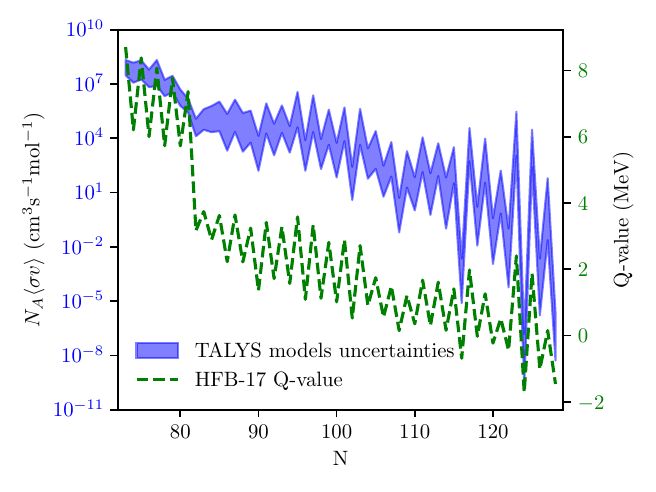}
\caption{The evolution of the neutron-capture rate $N_A \langle \sigma v \rangle$ evaluated at $T=1.0$~GK and the $(n,\gamma)$ $Q$-value along the Sb isotopic chain for all 48 theoretical TALYS models explained in Section~\ref{sec:nuclear_input}, calculated with the HFB-17~\cite{massmodel2} mass model 
(see text). \label{fig:isochain}}
\end{figure}

The statistical approach is a powerful tool when trying to estimate model prediction uncertainties, and it allows to study the impact of individual nuclear properties. 
However, it has two drawbacks. 
Firstly, by varying rates individually, one does not properly account for well-known nuclear properties such as pairing effects. 
It is well known that in general, neutron-odd nuclei will have a higher neutron-capture rate than neighboring neutron-even nuclei, as long as they are not close to a neutron magic number. 
In Fig.~\ref{fig:isochain}, using the Sb isotopic chain as an example, we see how the rates and $Q$-values gradually decrease the more neutrons are present in the nucleus. 
We also observe how odd-even effects (neutron pairing) lead to a higher capture rate and $Q$-value for isotopes with an odd number of neutrons, compared to their even neighbors.
Such correlations might get lost if the multiplicative factor is bigger than the typical difference between neighboring isotopes. 

Secondly, the statistical-approach simulations of the $r$ process are relying on one-zone models for the astrophysical conditions, meaning that the final abundances are reproduced by only one set of initial conditions. This is usually not the case in a real astrophysical scenario, where many zones with different initial conditions contribute to the final abundances. 
This means that while the variation of the neutron-capture rate in a nucleus in the statistical approach may lead to a significant change in the final abundances, this change might become insignificant when mixed together with the resulting abundances from the other ``zones''. 
In short, by disregarding the correlations between neighboring nuclei one may overestimate uncertainties in the final abundances, and the one-zone approach may point to single nuclei being important, while this importance could be averaged out when doing multi-zone calculations.

We label the second category of sensitivity studies the ``model-consistent approach''.
Here, the problems with the statistical approach are addressed by simulating the full astrophysical event where the $r$ process takes place (many trajectories with different physical parameters) and by employing the same nuclear models consistently for the whole nuclear chart. The nuclear uncertainties are propagated to the final abundances by repeating the simulation with different theoretical models of the quantity under consideration. 
This means that the final abundances represent the weighted sum of the abundances from different ``zones''. 
Furthermore, when it comes to neutron-capture rates, correlations between neighboring isotopes are accounted for as these are changed model-consistently. 
Although there are no studies for the $r$ process neutron-capture rates to date, one may find mass and $\beta$-decay sensitivity studies in the context of neutron star mergers by Kullmann \textit{et al.}~\cite{Kullmann2023} or for neutron-capture rates in the intermediate neutron-capture process by Goriely \textit{et al.}~\cite{Goriely2021}. 
However, also this approach comes with some drawbacks. 
First, potential errors and uncertainties in the theoretical models themselves are not accounted for (such as the choice of the interaction and its parameter uncertainties in a mean-field calculation), and this may lead to a significant underestimation of the actual uncertainties in the final abundances. 
Second, models may give reliable predictions of nuclei in a certain mass region, but not so in another, and this may lead to apparently confident but wrong abundance predictions.

In this work, we will investigate the impact of varying neutron-capture rates for the $r$ process using a model-consistent approach for five different trajectories. 
A trajectory represents the time evolution of density and temperature in the expanding ``bubble'' simulating material from e.g.\ a supernova or NSM. 
A single trajectory is equivalent to a one-zone simulation, and is therefore not necessarily representative for the actual nucleosynthesis outcome in the specific astrophysical sites. 
Nevertheless, the results can be compared to the predicted abundance uncertainties from the sensitivity studies using the statistical approach, and hopefully cast light on systematic and methodical sources of biases and errors in both methods. 
The details on the astrophysical simulations are described in Section~\ref{sec:Network_calc}, while the choice of nuclear inputs is discussed in Section~\ref{sec:nuclear_input}. In Section~\ref{sec:results} the results are shown and discussed, and a summary of the main findings is given in~\ref{sec:summary}.

\section{Network calculations}\label{sec:Network_calc}

In this work we make use of the five trajectories considered in Mumpower \textit{et al.}~\cite{Mumpower2016, *Mumpower2016e}. 
%Unless based on existing simulations, 
When not given directly by the multizone simulation,
the initial entropy $S$, the electron fraction $Y_e$ and the dynamical timescale $\tau$ for the trajectories are provided below:
\begin{itemize}
\item[a)] A ``hot'' $r$ process with low entropy: $S=30$ $k_B$, $Y_e = 0.20$, $\tau = 70$~ms.
\item[b)] A ``hot'' $r$ process with high entropy: $S=100$ $k_B$, $Y_e = 0.25$, $\tau = 80$~ms.
\item[c)] A ``cold'' $r$ process from a neutrino-driven wind (NDW) scenario modeled after Arcones \textit{et al.}~\cite{Arcones2007}, with artificially reduced $Y_e$ to 0.31.
\item[d)] A neutron-rich NSM trajectory modeled after Goriely \textit{et al.}~\cite{Goriely2011}.
\item[e)] A ``hot'' $r$ process with very high entropy: $S=200$ $k_B$, $Y_e = 0.30$, $\tau = 80$~ms.
\end{itemize}
Here, we have labeled the trajectory ``hot''   when the $(n,\gamma)\Leftrightarrow(\gamma,n)$ equilibrium is the dominant mechanism ($T\gtrsim 1$~GK until freezeout), and  ``cold'' when the trajectories  fall out of the $(n,\gamma)\Leftrightarrow(\gamma,n)$ equilibrium before the neutron flux is exhausted (see Fig.~1 in Mumpower et al.~\cite{Mumpower2016} for a plot of the temperature time evolution of the a) and c) trajectories).
These trajectories are used to run one-zone $r$-process simulations using SkyNet~\cite{Lippuner2017} for a time of 1~Gy. The reaction network includes electron screening and uses the JINA REACLIB library rates~\cite{JINAREACLIB} for all nuclear reactions except for neutron capture, as discussed in the following section. Spontaneous and neutron-induced fission was modeled with the rates of Panov \textit{et al.}~\cite{Panov2010}, with zero outgoing neutrons. Although we expect neutrons to be emitted during fission, results with two and four emitted neutrons per fission event did not yield appreciable differences in the final abundances for the region considered.

\section{Nuclear input}\label{sec:nuclear_input}

%In this work we study the variations in final abundance predictions by using different neutron-capture theoretical models. 
We calculate neutron-capture rates for every unstable neutron-rich nucleus using the nuclear reaction code TALYS 1.95~\cite{TALYS, TALYS2}. 
The code exploits the compound nucleus picture and uses the Wolfenstein-Hauser-Feshbach~\cite{Wolfenstein1951, HauserFeshbach1952} (WHB) formalism to calculate the neutron-capture rate. 
The main nuclear ingredients to the rate calculations are three nuclear statistical quantities: the optical-model potential (OMP), the nuclear level density (NLD) and the $\gamma$-ray strength function (GSF).

The neutron-capture rate $N_A\langle\sigma v\rangle(T)$ is calculated from the $(n,\gamma)$ cross section $\sigma_{n\gamma}$ as~\cite{ArnouldGoriely2007}
\begin{multline}\label{eq:MACS}
N_A\langle\sigma v\rangle(T)= N_A \left(\frac{8}{\pi \widetilde{m}}\right)^{1/2} \frac{1}{\left(k_B T\right)^{3/2}G(T)} \times \\ \int_0^\infty \sum_\mu \frac{2J_t^\mu + 1}{2J_t^0 + 1}\sigma^\mu_{n\gamma}(E)E\exp\left[-\frac{E+E_\mu}{k_B T}\right]\textrm{d}E,
\end{multline}
where $N_A$ is Avogadro's number, $\widetilde{m}$ the reduced mass of the target nucleus, $k_B$ Boltzmann's constant, $T$ the temperature, $J_t$ the target's spin, $E$ the relative energy between target and neutron, and $G(T)$ is the partition function defined as:
\begin{equation}\label{eq:partition_function}
	G(T) \equiv \sum_\mu \frac{(J_\mu + 1)}{(J_0 + 1)}e^{-E_\mu k_B}.
\end{equation}
The suffixes $\mu$ and $0$ represent the $\mu$-excited states and the ground state, respectively, and $\sigma^\mu_{n\gamma}$ represent thus the neutron-capture cross section when the nucleus is excited at a certain energy level $\mu$. 
%Generally, one can apply the Brink-Axel hypothesis~\cite{Brink, Axel} and state that the cross section built on an excited state is the same as if it were built on the ground state, meaning that $\sigma^\mu_{n\gamma} = \sigma_{n\gamma}$.
In the WHF formalism, the neutron-capture cross section is calculated as
\begin{equation}\label{eq:CN_ngamma}
	\sigma_{n,\gamma}(E_x) = \sum_{J,\pi} \sigma^\textrm{CN}_n(E_x,J,\pi) P_\gamma(E_x,J,\pi),
\end{equation}
where $\sigma^\textrm{CN}_n(E_x,J,\pi)$ is the probability that a free neutron ends up forming an excited compound nucleus of excitation energy $E_x$, spin $J$ and parity $\pi$, and $P_\gamma(E_x,J,\pi)$ the probability of the compound nucleus to $\gamma$-decay, effectively capturing the neutron. 
The first factor is described by the OMP, while the second can be expressed as
\begin{equation}\label{eq:P_gamma}
	P_\gamma(E_x,J,\pi) = \frac{\mathcal{T}_\gamma(E_x,J,\pi)}{\mathcal{T}_\textrm{tot}(E_x,J,\pi)},
\end{equation}
where $\mathcal{T}_\gamma(E_x,J,\pi)$ is the $\gamma$-transmission coefficient, and $\mathcal{T}_\textrm{tot} = \mathcal{T}_\gamma + \mathcal{T}_n$ is the total transmission coefficient (where $\gamma$-decay and neutron emission usually are the only two allowed decay channels).
Here, $\mathcal{T}_\gamma(E_x,J,\pi)$ may be expressed as
\begin{multline}\label{eq:Tgamma_f_rho}
	\mathcal{T}_\gamma(E_x,J,\pi) = 2\pi\sum_{X,L}\int_0^{E_x} E_\gamma^{2L+1} f^{XL}(E_\gamma) \times \\ \rho(E_x - E_\gamma, J, \pi)\textrm{d}E_\gamma,
\end{multline}
where $X$ and $L$ represent the electromagnetic mode (electric $E$ or magnetic $M$), $L$ the multipolarity, $E_\gamma$ the transition energy, and  $f$ and $\rho$ the GSF and the NLD, respectively.

TALYS 1.95 provides six different NLD models and eight GSF models. 
The NLD models with their respective TALYS keyword in parenthesis are:
\begin{itemize}
\item The constant-temperature + Fermi gas model (CTM, keyword \texttt{ldmodel~1})~\cite{GilbertCameron}
\item The Back-shifted Fermi gas model (BSFG, keyword \texttt{ldmodel~2})~\cite{GilbertCameron, VONEGIDY1988}
\item The Generalised Superfluid model (GSM, keyword \texttt{ldmodel~3})~\cite{ignatyuk79, ignatyuk93}
%\item The Skyrme-Hartree-Fock-Bogolyubov + statistical model (HFB+Stat, keyword \texttt{ldmodel~4}), tables from Ref.~\cite{GORIELY2001311}
\item The Hartree-Fock plus Bardeen-Cooper-Schrieffer statistical model (HFB+Stat, keyword \texttt{ldmodel~4}), tables from Ref.~\cite{DEMETRIOU200195}
\item The Hartree-Fock-Bogoliubov + combinatorial model (HFB+comb, keyword \texttt{ldmodel~5}), tables from Ref.~\cite{PhysRevC.78.064307}
\item The temperature-dependent Gogny-Hartree-Fock-Bogoliubov model (THFB+comb, keyword \texttt{ldmodel~6})~\cite{PhysRevC.86.064317}
\end{itemize}
where the first three are phenomenological models where fitting parameters are adjusted to reproduce known experimental data. The last three models are semi-microscopic, meaning that their predictions are based on a more fundamental treatment of the nuclear many-body problem.
However, we note that even these models are subject to adjustments through fit parameters in order for them to reproduce measured data such as known, discrete levels and $s$-wave neutron-resonance spacings.

\begin{figure*}[!]
\includegraphics[width=1.00\textwidth]{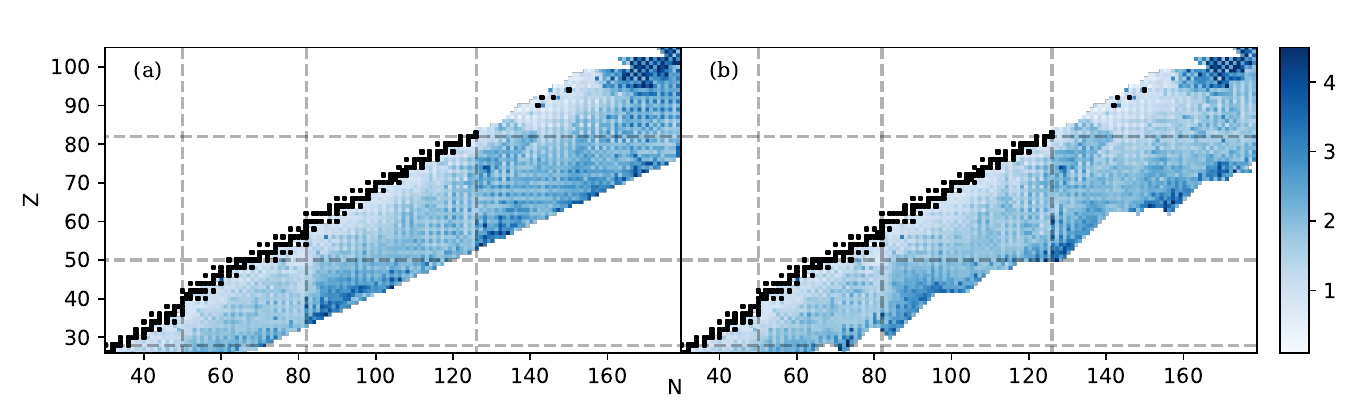}
\caption{(Color online) Differences (in orders of magnitude) between the highest and lowest predicted neutron-capture rate at $T = 1.0$ GK for all the 48 models using  (a) the FRDM-2012~\cite{FRDM2012} mass model (an updated version of what the JINA REACLIB rates~\cite{JINAREACLIB} are based on) and (b) the HFB-17~\cite{massmodel2} mass model. The plotted differences $\Delta(Z,N)$ have been calculated as $\Delta(Z,N) = \log_{10}(N_A\langle\sigma v \rangle_\textrm{max}) - \log_{10}(N_A\langle\sigma v \rangle_\textrm{min})$, where $N_A \langle\sigma v \rangle_\textrm{max,min}$ represent the maximum and minimum predicted rates, respectively. The different mass models predict different neutron drip lines, and this is the reason for the different shapes.
%We see how this is usually in the order of two to three orders of magnitude when it comes to nuclei involved in the $r$ process, with much higher differences for extremely neutron-rich nuclei.
\label{fig:extremes}}
\end{figure*}

\begin{figure}
\includegraphics[width=0.50\textwidth]{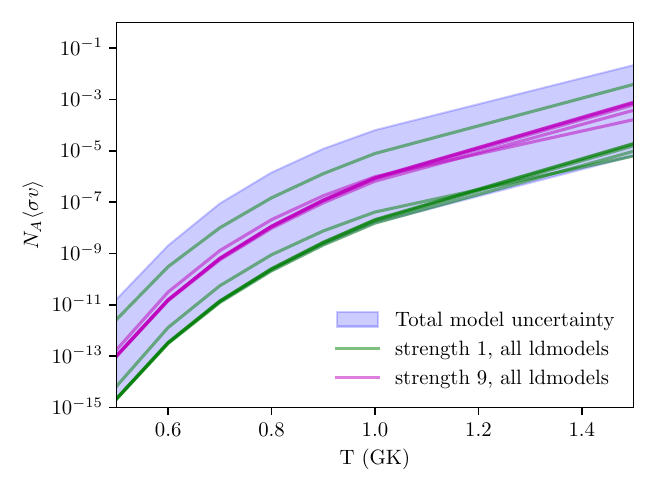}
\caption{(Color online) Neutron-capture rate predictions for different choices of GSF while varying the NLD models, using $^{182}$Xe as an exemplary case (see text).
\label{fig:NLD_GSF_effect}}
\end{figure}

\begin{figure}
\includegraphics[width=0.50\textwidth]{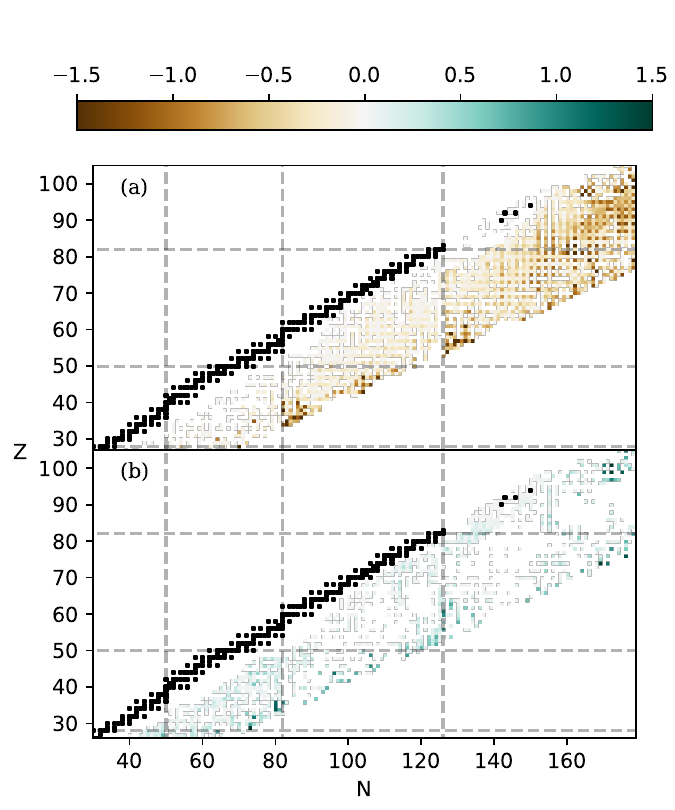}
\caption{(Color online) Difference between Figs.~\ref{fig:extremes}(b) and \ref{fig:extremes}(a), showing which mass model generates the biggest spread in predicted neutron capture rates, for all the 48 combinations of NLD and GSF models. The negative differences (i.e. where the FDRM-2012 model predicts larger uncertainties) are shown in the top panel, while the positive ones (where the HFB-17 predicts the larger uncertainties) in the bottom panel.
\label{fig:spread_diffs}}
\end{figure}

Correspondingly, the TALYS GSF models are:
\begin{itemize}
\item The Kopecky-Uhl generalized Lorentzian (GLO, keyword \texttt{strength~1})~\cite{KopeckyUhl1990}
\item The Brink-Axel standard Lorentzian (SLO, keyword \texttt{strength~2})~\cite{BRINK57,Axel}
\item The Hartree-Fock-BCS + QRPA tables based on the SLy4 interaction (SLy4+QRPA, keyword \texttt{strength~3})~\cite{GorielyKhan2002}
\item The HFB + QRPA calculation based on the BSk7 interaction (BSk7+QRPA, keyword \texttt{strength~4})~\cite{GorielyKhan2004}
\item The hybrid model (Hybrid, keyword \texttt{strength~5})~\cite{Goriely_hybrid_model}
\item The BSk7 + QRPA model with $T$-dependent width (BSk7T+QRPA, keyword \texttt{strength~6})~\cite{GorielyKhan2004}
\item The relativistic mean field + continuum QRPA calculation with $T$-dependent width (RMF+cQRPA, keyword \texttt{strength~7})~\cite{Daoutidis2012}
\item The Gogny-HFB + QRPA calculation complemented by low-energy enhancement (D1M+QRPA+0lim, keyword \texttt{strength~8})~\cite{Goriely2018}
\end{itemize}
where the first two again are phenomenological models, and the last six are semi-microscopic models.
From TALYS 1.96~\cite{TALYS196}, a ninth GSF model was introduced, namely the simplified modified Lorentzian (SMLO)~\cite{Goriely18b}. 
Although not included in the present study, this has now become the default choice of GSF model in TALYS. This exclusion is not expected to influence the results of the study, as the neutron-capture rate predictions using this model usually fall within the extremes of the other model combinations.

\begin{figure*}[!]
\includegraphics[width=1.00\textwidth]{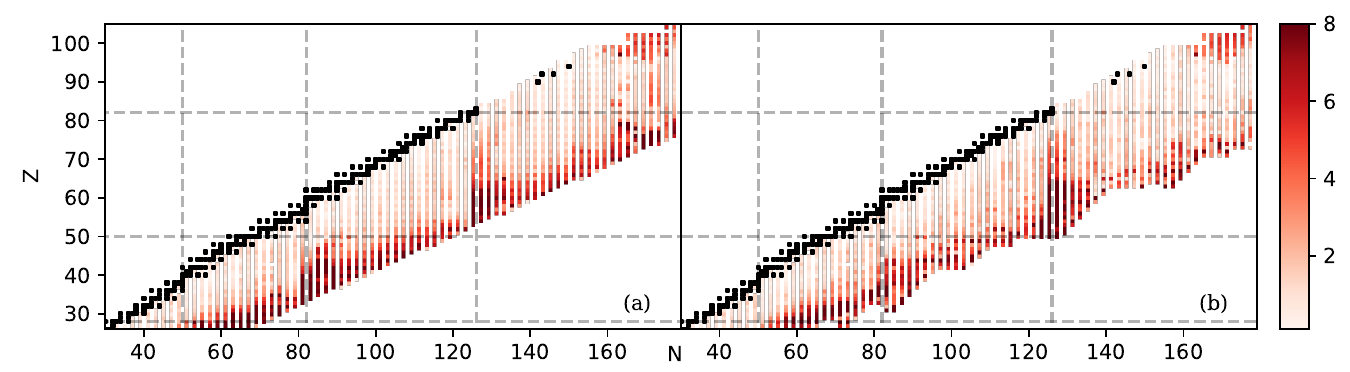}
\caption{(Color online) The biggest difference for each of the 48 models for the neutron-capture rate at $T=1.0$~GK of an $N$-odd nucleus $(Z,N)$ and its $N$-even neighbors $(Z,N-1)$ and $(Z,N+1)$. The difference (in orders of magnitude) is plotted against $(Z,N)$ for (a) the FRDM-2012~\cite{FRDM2012} mass model and (b) the HFB-17~\cite{massmodel2} mass model. The differences were calculated in a similar fashion as for those in Fig.~\ref{fig:extremes}.
%By measuring the ``biggest jump'' in neutron-capture rates between an N-odd isotope and its neighbors we can show the nuclide chart regions with the highest staggering effects.
\label{fig:staggering2}}
\end{figure*}

\begin{figure}
\includegraphics[width=0.50\textwidth]{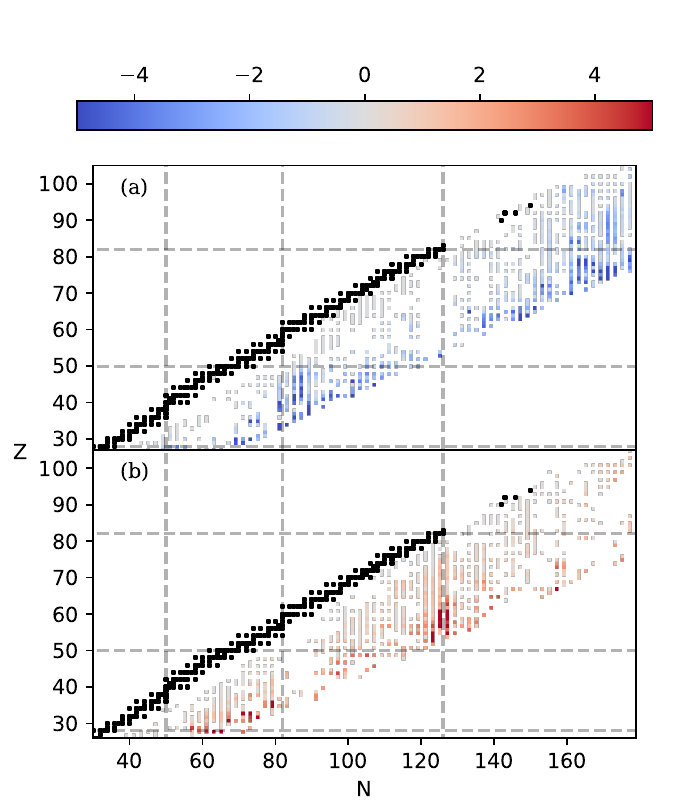}
\caption{(Color online) Difference between Figs.~\ref{fig:staggering2}(b) and \ref{fig:staggering2}(a), showing which mass model generates the highest staggering (see text). The negative differences are plotted in the top panel, while the positive ones in the bottom panel.
\label{fig:HFB17-FRDM}}
\end{figure}

While TALYS provides different OMP models, the choice of one model over another becomes gradually less significant when the temperature increases;  for typical $r$-process temperatures the OMP is considered to be of minor importance\footnote{This is not true if the isovector part of the potential is strongly enhanced, see Ref.~\cite{Goriely2007}.}. 
For this reason we only apply the Koning \& Delaroche OMP model~\cite{localomp}. 

Not part of the compound nucleus picture is the direct capture (DC) mechanism for neutron capture (see, e.g., Refs.~\cite{Oberhummer1991,Sieja2021} and references therein). 
This mechanism is expected to contribute with a small cross section, so that for many cases compound capture cross section completely dominates the total cross section. 
However, near the neutron drip line, close to and at neutron shell closures, the compound capture cross section becomes small enough so that the DC contribution may become appreciable and even the dominant part. 
We have chosen to not include DC this study, as its model predictions are rather uncertain, and we want to focus on the rates calculated via the compound nucleus picture, as these can be constrained with experimental data from Oslo-type experiments.

For the mass model, we use the FRDM-2012 model by M\"{o}ller \textit{et al.}~\cite{FRDM2012} (TALYS keyword \texttt{massmodel 1}) and the Skyrme-Hartree-Fock-Bogoliubov model~\cite{massmodel2} (HFB-17), corresponding to the TALYS keyword \texttt{massmodel 2}. Apart from being readily available in TALYS, these two mass models (or similar, updated versions of them) have been shown to produce comparable abundances in mass sensitivity studies, see e.g.\ \cite{Kullmann2023, Jiang2021}.

By combining the six NLD models and the eight GSF models we obtain 48 different neutron-capture rate models for each chosen mass model. 
These were calculated for all elements from Fe ($Z=26$) up to Sg ($Z=106$), from the first neutron-rich unstable isotope up to the drip line. Although the $Q$-value for neutron capture on $N$-even nuclei may already become negative before the drip line (see e.g.\ Fig.~\ref{fig:isochain}), the neutron-capture rates were nevertheless included for completeness, as  they can become relevant for trajectories with high neutron densities.

We note that TALYS provides different settings to modify and customize the input; nevertheless, we decided to keep these to the default ones, as the scope of the study is not necessarily to give an accurate description of the \textit{r} process, but to analyze the qualitative change in the predicted abundances when using the same neutron-capture rate model consistently throughout the nuclear chart. Starting from TALYS 1.96~\cite{TALYS196}, new default NLD and GSF models were introduced, where a new \texttt{strength 9} (the SMLO model introduced above) replaced the GLO as default GSF, and where $M1$ components such as a parametrized upbend and scissors mode were added to all GSFs. 
Although these structures in the GSF are observed experimentally for some nuclei, %and may impact the neutron-capture rate in an appreciable way, we still do not have enough data do predict their behavior well enough. 
there are still significant challenges in our understanding of these structures that make it highly questionable to add them to all nuclei on a general basis. 
For example, the scissors resonance is difficult to reproduce correctly, as models still overestimate or underestimate its strength when compared to experimental data (see e.g. Ref.~\cite{BelloGarrote2022}).
For the upbend, %the situation is made worse by the fact that the way it is modeled in TALYS is somewhat obscure, and 
we still do not have reliable systematics or even confirmed its presence throughout the nuclear chart even for stable nuclei, let alone exotic neutron-rich ones. 
It has so far mainly been observed in lighter nuclei (see e.g. Ref.~\cite{Midtbo2018}), and it is still not clear whether this feature is actually due to $M1$ transitions, or $E1$ transitions, or a mix of both~\cite{Jones2018}.
Nevertheless, there is no doubt that its presence could indeed impact the neutron-capture rates~\cite{Larsen_Goriely2010}.
We deem that a thorough study of the $M1$ scissors mode and the upbend would be highly important, but would also be outside the scope of the present study. 
Considering also that the SMLO model usually yields neutron-capture rates within the extremes produced by other model combinations, we decided that the use of TALYS 1.95 instead of 1.96 was both satisfactory with respect to the objective of the study, and gave us better control on the different models, parameters and features included.

All other nuclear reactions rates are described using the JINA REACLIB library~\cite{JINAREACLIB}. 
This library is based on the NON-SMOKER theoretical rates~\cite{RAUSCHERTHIELEMANN}. 
For every neutron-capture rate model we substitute the default JINA REACLIB neutron-capture rates (and those for their inverse reaction) with the TALYS-calculated ones. 
These, together with the default JINA REACLIB rates, are the 49 neutron-capture models used in the reaction network calculations in this work.

In Fig.~\ref{fig:extremes}, we show the differences between the highest and the lowest predicted neutron-capture rate out of the 48 TALYS calculations taken at temperature 1.0~GK in the astrophysical environment for the two mass models.\footnote{Because of a bug in TALYS, the rates using \texttt{strength 6} were not used for the isotopes of Tm ($Z=69$) between $A=215$ and $248$. For these few affected nuclei, the JINA REACLIB rates were used instead.}
We calculate this difference as 
\begin{equation}
\Delta(Z,N) = \log_{10}(N_A\langle\sigma v \rangle_\textrm{max}) - \log_{10}(N_A\langle\sigma v \rangle_\textrm{min})
\end{equation}
where $N_A \langle\sigma v \rangle_\textrm{max,min}$ represent the maximum and minimum predicted rates, respectively.
As can be seen, the deviations can be very large, in particular in the regions close to the neutron drip line and in the regions near the neutron magic numbers $N = 82,126$. 

It is unfortunately difficult to investigate if such large uncertainties are due to the deviations in NLD or GSF model predictions, since, as seen in Eq.~(\ref{eq:Tgamma_f_rho}), the two quantities are convoluted with each other in such a way that it is difficult to isolate the contribution of one of the two quantities. 
This means that, e.g., we could hold a GSF model constant, vary the six NLD models and obtain a large variation in the calculated neutron-capture rate predictions. 
This would lead us to believe that the choice of NLD model is responsible for such large uncertainties, until we try again by holding a different GSF model constant while varying the NLD models, and obtain more similar predictions for all employed GSF models. 
This effect is shown in Fig.~\ref{fig:NLD_GSF_effect}.
Even if some choices for the GSF model will yield a large variation in the predicted neutron-capture rates when changing the NLD model, others may instead yield rather similar predictions, here exemplified in the context of $^{182}$Xe($n,\gamma$) by using the GLO (\texttt{strength 1},~\cite{KopeckyUhl1990}) and the SMLO (\texttt{strength 9},~\cite{Goriely18b}) models for the GSF, respectively. 

Another consideration when using NLD and GSF models to calculate neutron-capture rates in the context of the \textit{r} process, is the fact that for very exotic nuclei with low (or even negative) ($n,\gamma$) Q-values, the neutron-capture cross section becomes very sensitive to levels (or rather, resonances) at very low excitation energies and their decay properties. 
However, the low-lying levels are experimentally unknown for the very neutron-rich region of the $r$ process. 
Moreover, the choice of bin size may also play a significant role for the neutron-capture rate calculation. 
When also considering that the NLD models are typically tested against neutron-resonance data at high excitation energies, it suggests that the level density at low $E_x$ in the exotic, neutron-rich mass region is particularly poorly constrained and, thus, the neutron-capture rate may be even more uncertain than what models predict.

We  note that there is a large difference in the rate predictions for very heavy nuclei around $Z = 100$ for both mass models. 
This might not be surprising as there is no or very little data in this region, due to the fusion-evaporation reactions used to produce very heavy and super-heavy nuclei that favor production channels with (multiple) neutron emission(s).
We also remark that the general trends of the two mass models applied here are very similar. 
Fig.~\ref{fig:extremes} shows that, for both mass models, the 48 predicted neutron capture rates tend to agree more when close to the valley of stability, than close to the drip line, as expected. In Fig.~\ref{fig:spread_diffs}, we show the difference between Figs.~\ref{fig:extremes}(b) and~\ref{fig:extremes}(a) as a way to quantify the difference in the predicted spread for the two mass models. 
Fig.~\ref{fig:spread_diffs}(a) shows for example that the FRDM-2012 overall generates larger uncertainties, especially for neutron-rich, $N> 126$ nuclei, and right after neutron shell closures for $Z\approx35, 55$ close to the drip line. Fig.~\ref{fig:spread_diffs}(b) shows that the HFB-17 model has larger uncertainties for light nuclei ($50<N<82$) and again near the drip line right before the $N=126$ isotone.

As a way to quantify the odd-even effects in each model, we would also like to investigate how the predicted neutron-capture rate changes when going from an $N$-odd nucleus $(Z,N)$ and its $N$-even neighbors $(Z,N-1)$ and $(Z,N+1)$.
This is shown in Fig.~\ref{fig:staggering2}, where the biggest difference between the neutron-capture rate of an $N$-odd nucleus and each of its neighbors is chosen and plotted. 
The differences are calculated in the same way as for Fig.~\ref{fig:extremes}, where a difference value of e.g.\ 6 means a 6-orders-of-magnitude difference in neutron-capture rate between an $N$-odd nucleus and one of its $N$-even neighbors, within the same model. Unsurprisingly, the biggest staggering is shown to be around $N$-shell closures, and near the drip line.

In order to investigate how the two mass models compare, we show in Fig.~\ref{fig:HFB17-FRDM} the difference between \ref{fig:staggering2}(b) and \ref{fig:staggering2}(a), again in orders of magnitude. The lower plot indicates where the HFB-17 mass model~\cite{massmodel2} generates the higher staggering in neutron-capture rates, and the upper plot for the FRDM mass model~\cite{FRDM2012}, correspondingly. 
We observe how the FRDM-2012 model (blue squares in Fig.~\ref{fig:HFB17-FRDM}(a)) predicts a larger staggering for neutron-rich nuclei around the $N=82$ isotone, and for the $N>160$ region, while the HFB-17 model (red squares in Fig.~\ref{fig:HFB17-FRDM}(b)) instead predicts larger staggering close to the $N=126$ isotone and generally for the $82<N<126$ region, again for neutron rich nuclei.

\section{Results}\label{sec:results}

\begin{figure}
\includegraphics[width=0.50\textwidth]{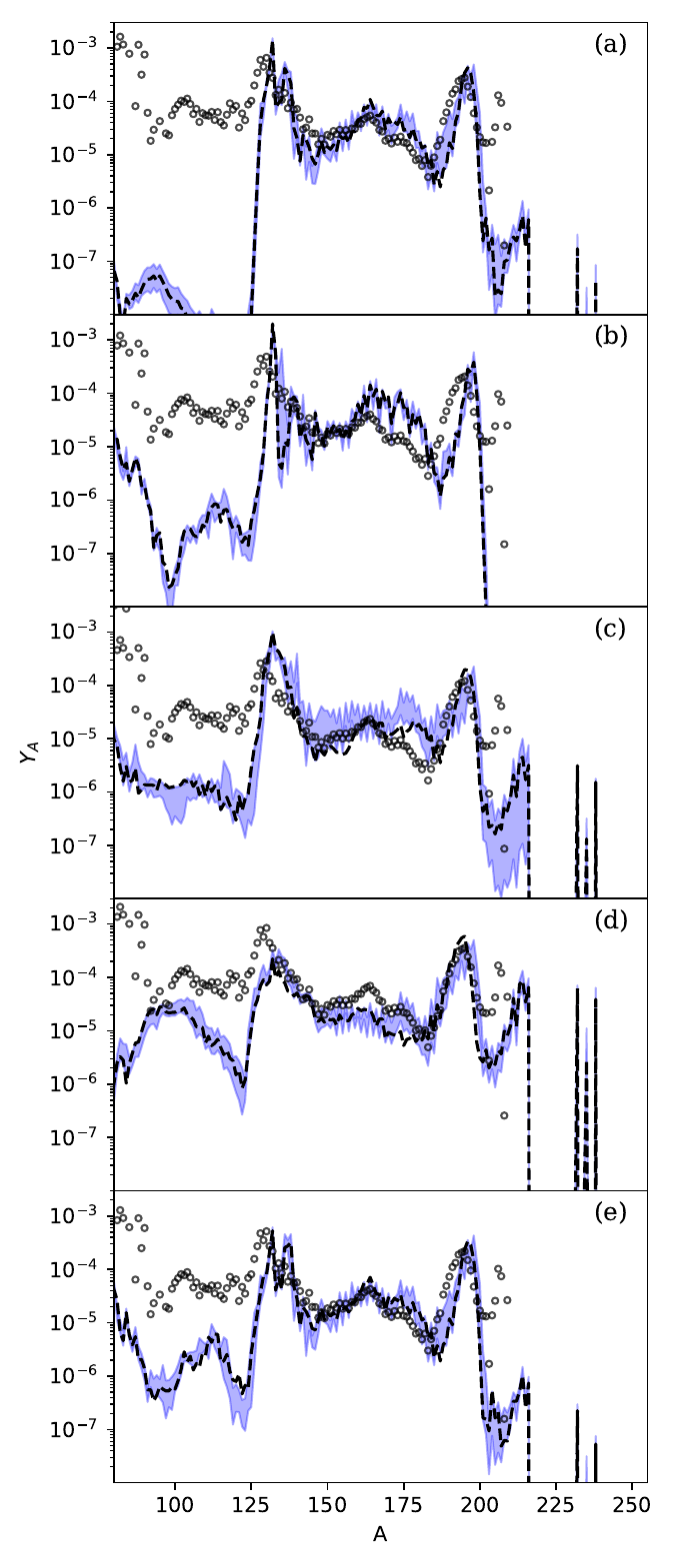}
\caption{The abundances from the five trajectories after 1~Gy evolution. In blue is the span of the final abundance predictions calculated from the 48 neutron-capture rate models. The black dashed line represents the abundances obtained using JINA REACLIB rates \cite{JINAREACLIB}, while the black dots are the $r$ solar abundances~\cite{Goriely1999} scaled down to fit the third peak at $A\approx 195$.\label{fig:abs}}
\end{figure}

The abundance yields obtained from the five trajectories using SkyNet can be seen in Fig.~\ref{fig:abs}\footnote{The data used for the figure can be found in Ref.~\cite{Abundance_dataset}.}.
We observe a marked staggering effect especially in the rare-earth peak. This may be mostly due to the choice of mass model, as the HFB-17 predicts strong odd-even effects in the nuclide chart region that would eventually $\beta$-decay to the rare-earth elements, see Fig.~\ref{fig:HFB17-FRDM}. 
This staggering in fact disappears for trajectories (c) and (d) when the JINA REACLIB rates are used, which are calculated with the NON-SMOKER formula~\cite{RAUSCHERTHIELEMANN}, using the FRDM mass model~\cite{FRDM2012}. 
Interestingly, the effect is preserved in all simulations for trajectory (e), including the one using JINA REACLIB. 
This is to be compared to the final abundances uncertainty band plotted in Fig.~11b from Mumpower \textit{et al.}~\cite{Mumpower2016, *Mumpower2016e}, where the same trajectory was employed with the same mass model (HFB-17), but by varying the neutron-capture rates using the statistical approach. 
Here the staggering is almost non-present, showing how the impact of shell effects and pairing correlations may indeed be washed out, if the assumed  model uncertainties are large enough to hide them.

Another noticeable result is the relatively small uncertainties we obtain  from using different neutron-capture models as compared to Fig.~11b in Mumpower \textit{et al.}~\cite{Mumpower2016, *Mumpower2016e}.
The same can be concluded when comparing to the uncertainty bands of Fig.~13 in Nikas \textit{et al.}~\cite{Nikas2020} for different trajectories. 
The uncertainties we obtain here are of course underestimated, because we have not considered parameter uncertainties  in the input models used to calculate the neutron-capture reaction rates. 
Even so, we find that the difference between the neutron-capture rate models used in this work may be of several orders of magnitude as shown in Fig.~\ref{fig:extremes}, similar to what the typical magnitude of the assumed rate errors are in the statistical approach.
This suggests that the inclusion of shell effects and pairing correlations does play  an important and non-negligible role in reducing prediction uncertainties in the calculated abundances.

Finally, in all the five different abundance yields in Fig.~\ref{fig:abs}, one can see how the $A\approx 186$ region has a bigger relative uncertainty compared to the rest. 
This is because it originates from the most neutron-rich region close to the $N=126$ isotone, above the (possibly) doubly-magic $^{176}$Sn and around the ``dark blue'' regions in Fig.~\ref{fig:extremes}(b). 
This region is extremely neutron-rich and exotic, with very large uncertainties in the predicted neutron-capture rates, which are again reflected in the final abundance predictions.

Unfortunately, the region around the $N=126$ isotones, is very difficult to study experimentally. 
However, although not reaching $^{176}$Sn, new and upcoming experimental facilities like FRIB~\cite{FRIB}, FAIR~\cite{FAIR} and the $N=126$ factory~\cite{ANL126factory}, might provide data in the region $Z=$70-80, $N\approx 126$, which would help constrain the theoretical mass models.
Moreover, methods for measuring neutron-capture cross sections in inverse kinematics as suggested by e.g.\ Reifarth \textit{et al.}~\cite{reifarth2017} and Dillmann \textit{et al.}~\cite{Dillmann2023} would be extremely valuable to get experimental information on  neutron-capture rates of neutron-rich nuclei.
Moreover, methods for measuring neutron-capture cross sections in inverse kinematics as suggested by e.g. Reifarth \textit{et al.}~\cite{reifarth2017} would be extremely valuable to get experimental information on  neutron-capture rates of neutron-rich nuclei.
Also indirect techniques like the Oslo method (Ref.~\cite{PPNP_larsen} and references therein), the surrogate method (see Ref.~\cite{escher2018} and references therein), the $\beta$-Oslo method~\cite{spyrou2014}, the shape method~\cite{Wiedeking2021,Mucher2023} and the inverse-kinematics Oslo method~\cite{ingeberg2020} would provide useful information to constrain the neutron-capture rates.

\section{Summary}\label{sec:summary}

The lack of confidence intervals in theoretical neutron-capture rates predictions poses a great problem for the correct quantification of final abundances errors of the $r$ process. 
This in turn makes the task of performing sensitivity studies difficult, and the conclusions on which nuclear properties have the biggest impact potentially questionable.
In this work we have discussed the strengths and weaknesses of two different approaches found in the literature, dubbed here as the ``statistical'' and the ``model-consistent'' approach. 
While the former tries to quantify the statistical errors in model predictions and attempts to propagate these using one-zone astrophysical models and interpret the results statistically, the latter sticks to one or few models and use them consistently for all the involved nuclei in more sophisticated, multi-zone simulations.

We have  presented the results of  network calculations using five different $r$-process trajectories, each having different inputs and representing different astrophysical scenarios. 
For each of these, 48 different neutron-capture rate models (plus the JINA REACLIB rates) were employed in order to estimate the sensitivity of the final abundances to these reaction rates. 
Although these are not meant to be interpreted as realistic representations of the $r$ process in these sites, this study provides some insights on the different strengths and weaknesses of the two approaches mentioned above. 

A staggering effect was observed especially for the rare-earth region in all trajectories, a feature that cannot be explained solely by the choice of mass model. 
The fact that this does not appear in similar studies using the statistical approach~\cite{Mumpower2016, *Mumpower2016e, Nikas2020}, even using the same mass model, suggests that the assumption of uncorrelated statistical errors in these studies may indeed mask the shell effects and pairing correlations, and probably overestimate the uncertainties. 
This is also corroborated by the fact that our uncertainties are markedly smaller than those obtained in the above mentioned studies, even though the rate uncertainties are of similar magnitude.

We do remark that the obtained uncertainties in this study are probably underestimated with regard to the real uncertainties, as uncertainties in the model parameters are not taken into account. A detailed  investigation including these uncertainties is beyond the scope of this work, but would be highly desirable to pursue in the future. 
We also note that experimental information for neutron-rich nuclei, especially near the $N=126$ closed shell, would be extremely valuable to better constrain the models in this mass region.

\section{Acknowledgments}

The calculations were performed on resources provided by Sigma2, the National Infrastructure for High Performance Computing and Data Storage in Norway (using ``Saga'' on Project No.~NN9464K).
The authors sincerely thank Matthew Mumpower for kindly providing the trajectories he applied in  his calculations, and Ina Kullmann and Stéphane Goriely for stimulating and enlightening discussions.
A.~C.~L. gratefully acknowledges funding from the Research Council of Norway, project grant no. 316116, and by the European Research Council through ERC-STG-2014 under grant agreement no. 637686. 

\bibliography{apssamp}% Produces the bibliography via BibTeX.

\end{document}